\documentclass[oupdraft]{bio}

\usepackage{amsmath}
\usepackage{amssymb}
\usepackage{graphicx}
\usepackage{mathrsfs}
\usepackage{multirow} % allows row spanning in tables

\usepackage{enumitem}
\setlist[itemize]{leftmargin=*}
\setlist[description]{leftmargin=*}

\usepackage{dsfont} % blackboard numerals

\usepackage{tikz}
\usetikzlibrary{matrix}
\usetikzlibrary{backgrounds}
\usetikzlibrary{positioning}
\usetikzlibrary{calc}
\usetikzlibrary{arrows}
\usetikzlibrary{shapes,patterns}
\usetikzlibrary{shapes.arrows}
\usetikzlibrary{decorations,decorations.pathmorphing}
\usetikzlibrary{decorations.pathreplacing}
\usetikzlibrary{decorations.shapes}
\usetikzlibrary{decorations.text}
\tikzset{
    every picture/.style={
        show background rectangle,
        inner frame sep=1pt,
        background rectangle/.style={
            draw=none
        }
    }
}

\usepackage[nolists,nomarkers]{endfloat}

\newcommand{\indicator}[1]{\mathds{1}\left\{ #1 \right\}}

\title{Markov counting models for correlated binary responses}
\author{FORREST W. CRAWFORD$^\ast$, DANIEL ZELTERMAN} % \\[4pt]
\begin{document}

%%%%%%%%%%%%%%%%%%%%%%%%%%%%%%%%%%%%%

\maketitle
\footnotetext{To whom correspondence should be addressed.}

\begin{abstract}{We propose a class of continuous-time Markov counting processes for analyzing correlated binary data and establish a correspondence between these models and sums of exchangeable Bernoulli random variables.  Our approach generalizes many previous models for correlated outcomes, admits easily interpretable parameterizations, allows different cluster sizes, and incorporates ascertainment bias in a natural way.  We demonstrate several new models for dependent outcomes and provide algorithms for computing maximum likelihood estimates.  We show how to incorporate cluster-specific covariates in a regression setting and demonstrate improved fits to well-known datasets from familial disease epidemiology and developmental toxicology.}{
% Keywords
Markov process; 
Bernoulli trials;
developmental toxicity;
familial disease;
teratology.
}
\end{abstract}

%%%%%%%%%%%%%%%%%%%%%%%%%%%%%%%%%%%%%

\section{Introduction}

The simplest statistical model for a collection of $n$ binary outcomes is the binomial distribution, which assumes that responses are independent and identically distributed.
%\citep[][page 19]{Collett2002Modelling}.  
However, many investigations have found that the binomial distribution sometimes gives a poor fit to certain types of data \citep{Greenwood1920Inquiry,Haseman1976Distribution,Altham1978Two}.  This empirical observation, along with suspicions that the mechanism generating the outcomes might induce dependencies, has encouraged development of more flexible models that account for correlations in responses.  Dependent or correlated binary data arise commonly in studies of developmental toxicology and litter size \citep{Williams1975Analysis,Kupper1978Use,Altham1978Two}, familial disease aggregation \citep{Liang1992Multivariate,Yu2002Statistical}, or when ascertainment considerations necessitate a biased approach to sampling \citep{Matthews2008Analysis}.  Groups of dependent responses are often called ``clusters'', and in many applications the response of interest is the number of affected units in a cluster with $n$ members.  

%As an illustrative example, Table \ref{tab:ipfdata}  shows the observed frequencies of 60 cases of interstitial pulmonary fibrosis (IPF) in the siblings of families with at least one case of chronic obstructive pulmonary disease (COPD).  The triangular shape of Table \ref{tab:ipfdata} arises because the family/cluster size limits the number of individuals that can be affected.  The simplicity of the dataset belies its complexity; the binomial model provides a poor fit and researchers have proposed a variety of remedies for the dependency of outcomes within clusters.
%\begin{table}
  %\centering
  %\begin{tabular}{llccccccc}
%\hline
%Number of & Number of & \multicolumn{7}{l}{Number of } \\
 %siblings & families & \multicolumn{7}{l}{affected siblings }\\
%\hline
%$n$ &     & 0 & 1 & 2 & 3 & 4 & 5 & 6 \\
%\hline
%1 & 48 & 36 & 12 \\
%2 & 23 & 15 & 7 & 1 \\
%3 & 17 & 5  & 7 & 3 & 2 \\
%4 & 7  & 3  & 3 & 1 & 0 & 0 \\
%6 & 5  & 1  & 0 & 1 & 1 & 1 & 0 & 1 \\
%\hline
  %\end{tabular}
  %\caption{Observed frequencies of interstitial pulmonary fibrosis (IPF) in siblings of patients with chronic obstructive pulmonary disease (COPD), from a dataset analyzed by \citet{Liang1992Multivariate} and later by \citet{Yu2002Statistical}.  The binomial distribution provides a poor fit to these data, possibly indicating a household or genetic component to disease risk.}
  %\label{tab:ipfdata}
%\end{table}

When individual unit-level data are available, mixed-effects logistic regression approaches \citep[e.g.][]{Stiratelli1984Random} can model correlation using cluster-specific effects; marginal models posit a population-averaged mean and a working covariance structure \citep{Zeger1986Longitudinal}.  These approaches depend on the access to individual-level outcomes, which is not always available.  Mixed-effects and marginal models allow specification of pairwise covariances, but may be unable to provide higher-order dependency between outcomes.  This has led researchers to study models for the sum of dependent Bernoulli variables. %\citep{Bahadur1959Representation,Williams1975Analysis,Kupper1978Use,Altham1978Two,Witt2013Simple}.  
One of the simplest is the beta-binomial model, used to account for extra-binomial variation in clustered counts \citep{Moore2001Exploring,Yu2002Sums}.  
%Researchers have also sought to model the correlations in the outcomes directly.  
\citet{George1995Full} and \citet{Bowman1995Saturated} present general expressions for the likelihood of a sum of exchangeable Bernoulli variables via a combinatorial argument.
%using joint probabilities of outcomes for subsets of responses.  
In this context, exchangeability means that the joint probability of all the outcomes in a cluster is invariant to permutation of the responses, a notion we define more formally in Section \ref{sec:sums}.  \citet{Kuk2004Litter} uses the \citet{George1995Full} framework to define families of power functions that show superior fit in developmental toxicity studies and \citet{Pang2005Shared} give a model that allows a random subset of responses to share their response.  
%\citet{Yu2001Exact,Yu2002Statistical} describe new dependency models and a framework for exact testing.  
\citet{Yu2002Sums,Yu2008Sums} derive the beta-binomial distribution and other models under the \citet{George1995Full} framework.  Several authors describe methods to fit data consisting of observations on clusters of different sizes: \citet{Stefanescu2003Likelihood} interpret different cluster sizes in a missing data framework and derive EM algorithms for fitting.  \citet{Xu2003Modelling} and \citet{Pang2007Test} deal with this issue by assuming that the marginal distributions of the first $k$ responses in different cluster sizes are equal.  

In this work, we take a very different approach: we show that sums of exchangeable Bernoulli random variables can be represented as continuous-time Markov counting processes via a technique called probabilistic embedding \citep{Blom1991Embedding}.
%(\citealp{Blom1991Embedding}; \citealp[][page 186]{Blom1994Problems}).  
By introducing an auxiliary variable, the binary responses are made to depend on the arrival times of points in a Markov counting process.  
%Markov counting models for dependent counts are also known as ``generalized Poisson'' or ``pure birth'' processes \citep[][page 119]{Karlin1975First}.  
This formulation provides a flexible way to parameterize and fit models of correlated binary outcomes, and accommodates different cluster sizes and ascertainment schemes. 
%Markov processes are also easy to parameterize so that estimated parameters have intelligible units.  
We review basic results for exchangeable Bernoulli variables and give examples of models derived under this framework.  We then describe a class of Markov counting process and give five examples inspired by principles from infectious disease epidemiology.  Next, we show that any Markov counting process can be expressed as a sum of exchangeable Bernoulli variables.  
%We discuss unequal cluster sizes and ascertainment bias.  
We apply our approach to three datasets in which outcomes cluster in families and one developmental toxicology experiment.  Supplementary Appendices provide simulation results, algorithms for maximum likelihood estimation, regression with covariates, and numerical evaluation of likelihoods.

%%%%%%%%%%%%%%%%%%%%%%%%%%%%%%%%%%%%%%%%%%%%%%%%

\section{Sums of exchangeable Bernoulli variables}

\label{sec:bg}
\label{sec:sums}

\citet{George1995Full} and \citet{Bowman1995Saturated} describe a likelihood framework for sums of exchangeable Bernoulli random variables that depends on knowledge of joint probabilities of subsets of variables taking value 1.  Consider a sequence of $n$ exchangeable Bernoulli variables $Z_1, \ldots, Z_n$.  By exchangeability, we mean that the joint probability of a collection of variables taking certain values is invariant to reordering.  More formally,
  $\Pr(Z_1=z_1,\ldots,Z_n=z_n) = \Pr\left(Z_{\pi(1)}=z_{\pi(1)},\ldots,Z_{\pi(n)}=z_{\pi(n)}\right)$
for any permutation $\pi$ of the indices $1,2,\ldots,n$ \citep{deFinetti1931Funzione}.  Now consider the probability that $r$ of the $Z_i$'s take value 1 and $n-r$ take value 0.  By exchangeability, we can express this as the joint probability that the first $r$ take value 1 and the remainder are 0.  Let $\lambda_j = \Pr(Z_{i_1}=Z_{i_2}=\cdots=Z_{i_j}=1)$ be the joint probability that every $Z_i$ for $i\in I_j$ is 1, where the cardinality of the set $I_j$ is $j$.  
%Then
%\begin{equation}
  %\begin{split}
    %\Pr(Z_1=\cdots=Z_r=1,Z_{r+1}=\cdots=Z_n=0) %&= \sum_{j=0}^{n-r} (-1)^j \binom{n-r}{j} \Pr(Z_1+\cdots+Z_r=r,\ Z_{r+1}+\cdots+Z_n=j) \\
        %= \sum_{j=0}^{n-r} (-1)^j \binom{n-r}{j} \lambda_{r+j}
  %\end{split}
 %\label{eq:atleastr}
%\end{equation}
%using the inclusion-exclusion formula for the probability that \emph{at least} $j$ of $Z_{r+1},\ldots,Z_n$ are 1.  
Now letting $Y_n=\sum_{i=1}^n Z_i$, application of the inclusion-exclusion formula gives
\begin{equation}
 \Pr(Y_n=r) = \binom{n}{r}\sum_{j=0}^{n-r} (-1)^j \binom{n-r}{j} \lambda_{r+j} .
 \label{eq:PrYn}
\end{equation}
A derivation of \eqref{eq:PrYn} is given by \citet[][page 513]{George1995Full}.  By specifying the joint probabilities $\lambda_j$ for $j=0,\ldots,n$, the distribution of any sum of exchangeable Bernoulli variables can be represented.  In particular, setting $\lambda_{r+j}=p^{r+j}$ recovers the binomial distribution.  The $\lambda_j$'s are sometimes called ``marginal'' probabilities \citep{Dang2009Unified}, since they express the joint probability of $j$ successes, summed over all possible outcomes of the remaining $n-j$ variables.  This model is called ``saturated'' when all the $\lambda_j$'s are allowed to be nonzero.
%, resulting in a dependency model with correlations of all orders between outcomes \citep{George1995Full}.  
%A saturated model of clustered data, where each cluster has the same size $n$, is nonparametric in the sense that it does not rely on a particular parameterization of the $\lambda_j$'s.

We note three major issues with the model of \citet{George1995Full} given by \eqref{eq:PrYn}.  First, it is unclear how to interpret the joint probabilities $\lambda_j$ or correlations when analyzing data from clusters of different sizes since the number of unknown parameters for each observation is equal to the cluster size.  
%For example, the saturated model for a cluster of size $n$ depends on $\lambda_1,\ldots,\lambda_n$, and the model for a cluster of size $n'>n$ depends on $\lambda_1,\ldots,\lambda_n,\ldots,\lambda_{n'}$, and these probabilities may not be the same for $j=1,\ldots,n$.  
\citet{Xu2003Modelling} and \citet{Pang2007Test} deal with this problem by assuming that the marginal probability of $r$ responses having value 1 in a family of size $n\geq r$ is equal to the probability of $r$ responses having value 1 in a family of size $n'>n$, but this assumes response probabilities do not depend cluster size.  Second, it can be difficult to specify joint probabilities $\lambda_j$ for $j=0,\ldots,n$ that result in a well-defined probability mass function \citep{George1995Full,Stefanescu2003Likelihood}.
%give monotonicity constraints on the $\lambda_j$'s and \citet{Stefanescu2003Likelihood} provide diagrams of their feasible regions.  
Often one must solve a non-trivial combinatorial problem in order to specify the $\lambda_j$'s \citep[see, e.g.][]{Kuk2004Litter,Pang2005Shared}.  
%The situation becomes more complicated in regression settings: \citet{Dang2009Unified} define families of binomial mixtures using completely monotonic functions to satisfy these conditions.  
Third, sampling or ascertainment of clusters can sometimes depend on the responses; for example, often families in epidemiological studies are selected via a single affected member.  The likelihood of observing $r$ affected individuals in a family of size $n$ must then be computed conditional on having at least one response having value 1, which may be a function of family size $n$.  The interaction of ascertainment conditions and varying cluster sizes can substantially complicate inference for dependent counts.

\subsection{Examples of models for $\lambda_j$}

\subsubsection{Binomial}

When the Bernoulli variables are independent with probability $p$ of success, $\lambda_{j}=p^j$ and \eqref{eq:PrYn} reduces to the binomial probability $\Pr(Y_n=r) = \binom{n}{r} p^r (1-p)^{n-r}$.

\subsubsection{Beta binomial}

\citet{Yu2008Sums} show that setting $\lambda_1=p$, $\lambda_2 = p(p+\alpha)/(1+\alpha)$, and $\lambda_k = p(p+\alpha)\cdots(p+(k-1)\alpha)/(1+\alpha)\cdots(1+(k-1)\alpha)$ for $3 \leq k\leq n$ gives the beta binomial distribution 
\[ \Pr(Y_n=r) = \binom{n}{r} \left. \prod_{k=0}^{r-1} (k\alpha+p) \prod_{s=0}^{n-r-1} (s\alpha+1-p) \middle/ \prod_{j=0}^{n-1} (j\alpha+1) \right. \]
when $\alpha > -\min\{ p,1-p\}/(n-1)$.  Here, $p$ is the marginal success probability, and $\alpha$ is a measure of correlation.  Setting $\alpha=0$ recovers the binomial distribution. 

\subsubsection{$q$-power}

Consider the family of distributions in which $\lambda_j = p^{j^\gamma}$, where $p$ is called the marginal response probability.  When $0\leq\gamma\leq 1$, the probability distribution \eqref{eq:PrYn} is well-defined.  \citet{Kuk2004Litter} proposes to set $q=1-p$ and model the number of zero outcomes, $n-Y_n$.  Then \eqref{eq:PrYn} becomes
$\Pr(n-Y_n=r)=\binom{n}{r}\sum_{j=1}^r (-1)^j \binom{r}{k} q^{(n-r+j)^\gamma}$. 
Here, $\gamma$ is a measure of positive intra-cluster correlation: setting $\gamma=1$ results in no correlation between responses.

%%%%%%%%%%%%%%%%%%%%%%%%%%%%%%%%%%%%%%%%%%%%%%%%

\section{Markov counting processes}

\label{sec:counting}

There is an important correspondence between the \citet{George1995Full} representation \eqref{eq:PrYn} and continuous-time Markov counting models.  To make this clear, we formally define this class of processes and show how to calculate their transition probabilities.  In the next Section, we construct an equivalence between Markov counting processes and sums of exchangeable Bernoulli random variables.  Consider a continuous-time Markov process $X(t)$ that counts the number of arrivals (or points) before time $t$.  When $k$ points have arrived, the rate of arrival of the next point is $\mu_k$.  Let $P_{mr}(t) = \Pr(X(t)=r\mid X(0)=m)$ be the probability that at time $t$ there have been $r$ arrivals, given that there were $m$ already at time $0$.  This probability obeys the forward equation
\begin{equation}
\frac{\text{d}P_{mr}(t)}{\text{d}t} = \mu_{r-1} P_{m,r-1}(t) - \mu_r P_{mr}(t) 
\label{eq:odes}
\end{equation}
where $\mu_r>0$ is the instantaneous rate of the $r+1$st arrival, given that $r$ have already arrived \citep[][page 119]{Karlin1975First}.  This counting model is also known as the ``generalized Yule'' or ``pure birth'' process.  The homogeneous Poisson process with $\mu_r=\mu$ is the best-known counting process, with transition probability $P_{mr}(t) = (\mu t)^{r-m} e^{-\mu t}/(r-m)!$.  For a general Markov counting process with rates $\mu_k$, $k=0,1,\ldots$, the transition probability is
\begin{equation}
P_{mr}(t) = \left(\prod_{k=m}^{r-1}\mu_k\right) \sum_{k=m}^r \left(\prod_{\ell\neq k}(\mu_\ell-\mu_k)\right)^{-1} \exp[-\mu_k t] 
\label{eq:purebirth}
\end{equation}
for $0\le m \le r$ and $t>0$ when $\mu_k \neq \mu_r$ for all $m$ and $r$ \citep[][page 65]{Renshaw2011Stochastic}.  For a given set of rates $\{\mu_k\}$, simpler representations of the likelihood \eqref{eq:purebirth} are often available, as we show in section \ref{sec:examples}.  When $\mu_k=\mu_\ell$ for some $k$ and $\ell$, it can be more difficult to derive likelihood expressions. Fortunately, computational evaluation of the likelihood is straightforward and robust via numerical methods.  We give a general method for numerically evaluating $P_{mr}(t)$ in the Supplementary Appendix.

\subsection{Examples of models for $\mu_j$}

\label{sec:examples}

It can be challenging to translate informal ideas about dependency into parametric models for dependent count data in the framework of \citet{George1995Full}.  However, counting process rates are often easy to specify; usually a consideration of the conditional risk of a new event, given the number that have already occurred, is enough to express the $\mu_k$'s in a useful form.  
%Usually every $\mu_k$ is a function of cluster size $n$; since $\mu_n=0$, probability is conserved on $\{0,\ldots,n\}$.  
Modelers do not need to accommodate awkward constraints on the rates, such as monotonicity, that might make them difficult to specify jointly or interpret \citep[see][for example]{Stefanescu2003Likelihood}.  
%Once $\mu_k$, $k=0,\ldots,n-1$ are specified, inference is straightforward either by numerical maximization of the likelihood or the EM algorithm approach for the marginal likelihoods outlined in Appendix \ref{app:ml}.
%To illustrate the ease of modeling using counting processes, 
Here we present five simple counting processes derived from basic principles of infectious disease epidemiology. We imagine a household of size $n$ with $k$ members already affected by the disease.
%The number of individuals in the household who currently have the disease is $k$, %$Y_n = \sum_{i=1}^n Z_i$, 
%Family members' outcomes may be dependent because affected members may transmit the disease to unaffected members.  
Transmissibility of disease status induces dependency in the outcomes of individual family members; households are ``clusters'' and individuals are ``units''.  We distinguish between two sources of risk to members of a cluster of size $n$: exogenous or extra-cluster risk to which all unaffected units are subject, and infectivity, or risk experienced by each susceptible member in proportion to the number already affected.
%The risk experienced by the cluster due to exogenous factors is proportional to the number of unaffected units, since they are \emph{susceptible} to becoming affected:
%$\text{exogenous incidence} = (\text{exogenous risk}) \times (\text{\# susceptible})$ .
%The risk to the cluster therefore decreases as the number of affected units increases.  In the household disease example, exogenous risk is the constant chance of infection from the world outside the household.  
%$ \text{within-cluster incidence} = (\text{infectivity})\times (\text{\# affected}) \times (\text{\# susceptible})$ .
%In an infectious disease context, the intensity of this risk per susceptible-affected pair is called infectivity, contact rate, or secondary attack rate \citep{Britton1997Tests}.  The within-cluster risk is smallest when either 0 or $n$ members of the cluster are affected, and it is largest when a roughly equal number of members are affected and susceptible.  
%%Models combining both of these types of risk are similar to susceptible-infective (SI) models in epidemiology \citep{Andersson2000Stochastic}.
Table \ref{tab:examples} shows a summary of the counting processes we consider in what follows.  %the rate function $\mu_k$, an example realization of the counting process, and a schematic diagram of the risk for each of the models introduced below.  In the example realization, vertical lines indicate the sequence of arrivals and the gray curve shows the arrival rate function $\mu_k$ for each interval on which $X(\tau)=k$.  The number of arrivals between time 0 and $t$ is $X(t)$.  %For each model, we specify the arrival rates $\mu_k$ for $k=0,\ldots,n-1$. % and whenever possible, we give an expression for the likelihood $P_{mr}$.  
%The Supplementary Material provides the derivation for each model likelihood.  

\begin{table}
  \centering
  \begin{tabular}{llll}
    Model & Rate $\mu_k$ & Counting process example & Risk schematic \\
    \hline
    Susceptible-1
    &
    $ \alpha(n-k)$
    & 
    \parbox[c]{1em}{ \includegraphics[width=1.6in]{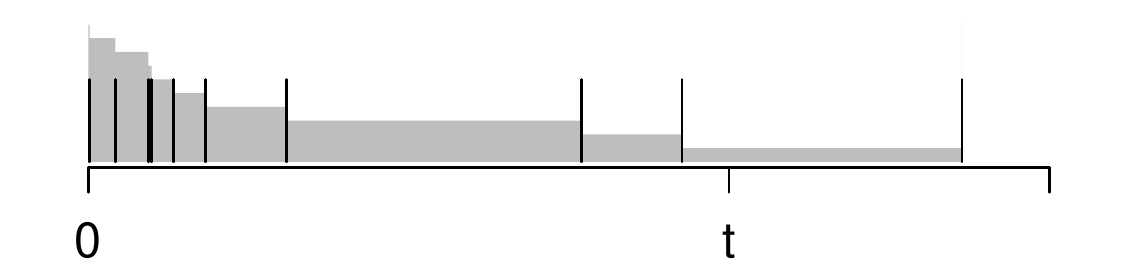} }
    &
    \multirow{2}{*}{ \parbox[c]{1em}{ \includegraphics[width=1.2in]{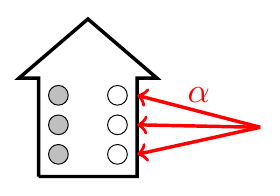} } } \\[3em]
Susceptible-2
&
    $ \alpha(n-k)^\gamma$
    & 
    \parbox[c]{1em}{ \includegraphics[width=1.6in]{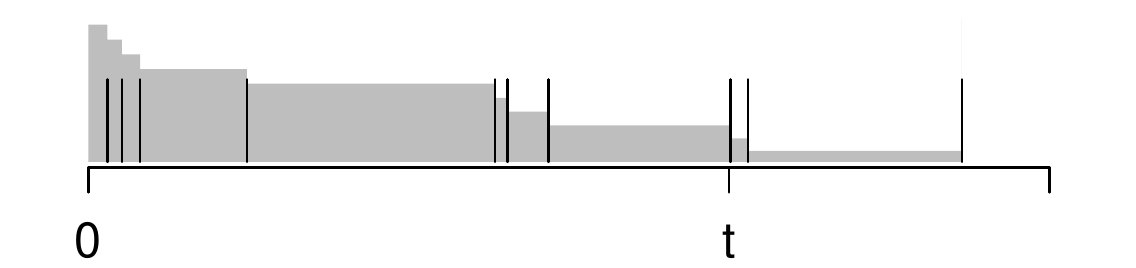} }
    &
\\
\hline
Infectivity-1
&
$ \beta k (n-k)$ 
& 
    \parbox[c]{1em}{ \includegraphics[width=1.6in]{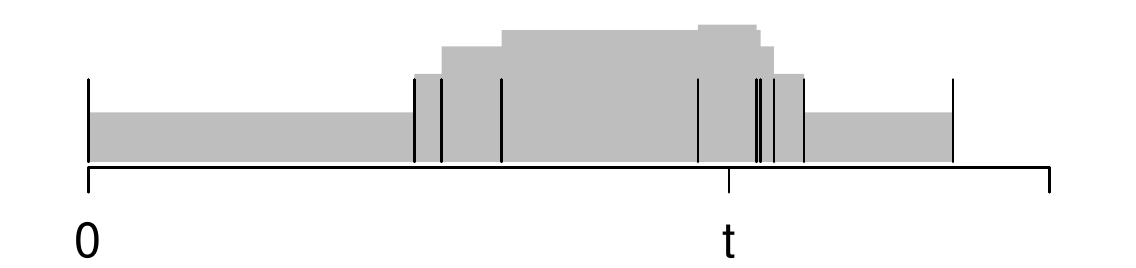} }
&
\multirow{2}{*}{\parbox[c]{1em}{ \includegraphics[width=0.77in]{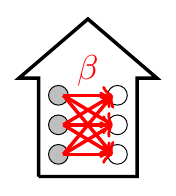} } } \\[3em]
Infectivity-2
&
$ \beta k^\eta (n-k)^\gamma$ 
& 
    \parbox[c]{1em}{ \includegraphics[width=1.6in]{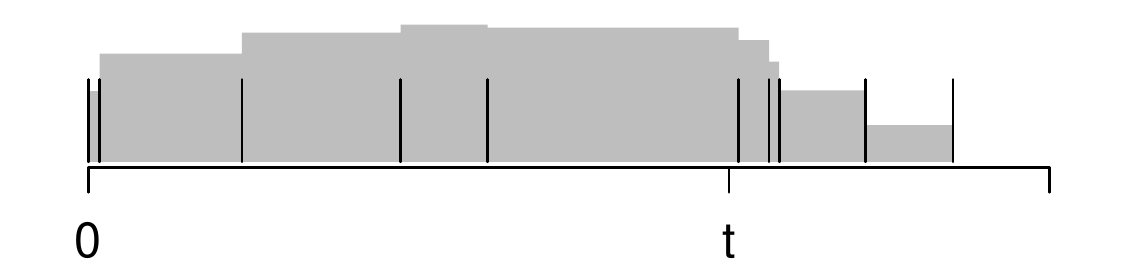} }
&
\\
\hline
Combined
&
$ \alpha(n-k) + \beta k(n-k)$
& 
    \parbox[c]{1em}{ \includegraphics[width=1.6in]{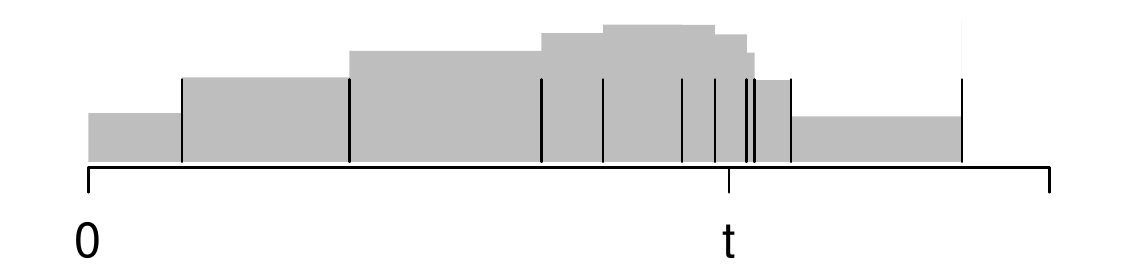} }
&
    \parbox[c]{1em}{ \includegraphics[width=1.2in]{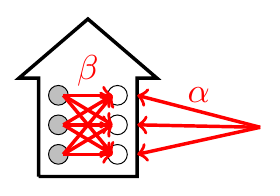} }
\\

\hline
\end{tabular}
\caption{Illustration of the proposed counting process models. Model name and arrival rate $\mu_k$ are given in the first two columns.  A stochastic realization of the counting process is shown, where a vertical line represents the time of each arrival and the gray step function represents the rate $\mu_k$.  A schematic diagram of a household is given for each type of model.  Filled gray circles represent affected family members and white circles represent unaffected members; in each diagram there are $n=6$ family members with 3 affected and 3 unaffected.  Exogenous or extra-household risk per unaffected member is $\alpha$, and the risk per potential contact between affected and unaffected members is $\beta$.}
\label{tab:examples}
\end{table}

%%%%%%%%%%%%%%%%%%%%%%%%%%%%%%%%%%%%%%%%%%%%%%%%%%%%

\subsubsection{Susceptible}
\label{sec:susc}

Consider a cluster of size $n$ in which each unaffected (susceptible) unit experiences the same exogenous risk $\alpha>0$.  When there are $k$ affected units, the number of unaffected units is $n-k$ and the risk to the cluster is $\mu_k = \alpha(n-k)$.  This formulation produces a counting process with a familiar epidemiological interpretation corresponding to constant per-unaffected-unit risk and no infectivity between units.  In fact, this model is formally equivalent to the binomial model with success probability $1-e^{-\alpha}$.  The likelihood for this ``susceptible-1'' model is 
$P_{mr} = \binom{n-m}{r-m} e^{-\alpha(n-r)} (1-e^{-\alpha})^{r-m}$.
We report this fact here to show that the susceptible counting process model, which has a traditional epidemiological interpretation, corresponds exactly to the simplest model for $n$ binary outcomes.  One straightforward extension of the susceptible-1 model is to allow the cluster risk to be a non-negative power function of the number of susceptibles, $\mu_k = \alpha(n-k)^\gamma$, where $\alpha>0$ and $\gamma>0$.  If $0<\gamma<1$, the cluster experiences risk smaller than that obtained by the susceptible-1 model, and if $\gamma>1$, the cluster experiences greater risk.  
%In this case, the likelihood for the susceptible-2 model is
%\[ P_{mr} = \frac{(n-m)!^\gamma}{(r-m)!^\gamma} \sum_{j=m}^r \frac{e^{-\alpha(n-r)^\gamma}}{\prod_{\ell\neq j} [(n-\ell)^\gamma - (n-k)^\gamma]}. \]

\subsubsection{Infectivity}

In contrast to the susceptible models, the infectivity-1 model considers only risk due to affected cluster members.  Each potential contact between susceptible and affected units presents an opportunity for a new case.  When there are $k$ affected units, the number of ways one affected and one susceptible unit can come into contact is $k(n-k)$, so $\mu_k = \beta k(n-k)$ where $\beta>0$ is the per-contact infectivity. This model formalizes the epidemiological notion of infectivity or contagion in a closed community \citep{Britton1997Tests}.  
%In the infectivity-1 model, the arrival rates are symmetric: $\mu_k = \mu_{n-k}$ so the likelihood cannot be evaluated via \eqref{eq:purebirth}, and we resort to numerical methods.
%, outlined in the Supplementary Materials.  
Since $\mu_0=0$, this model is most useful when ascertainment is of clusters with at least one affected member.  The infectivity-2 model extends the infectivity-1 model to allow the cluster risk to vary as a power of the number of affected and susceptible members, $\mu_k = \beta k^\eta (n-k)^\gamma$, where $\beta$, $\eta$, and $\gamma$ are non-negative.

\subsubsection{Combined}
\label{sec:si}

Now we combine the susceptible-1 model with the infectivity-1 model.  The per-susceptible risk from extra-cluster sources is $\alpha$, and the risk contributed by one affected member to each susceptible is $\beta$.  %Informally, 
%\begin{equation*}
  %\text{rate} = (\text{exogenous risk})\times(\text{\# susceptible}) + (\text{infectivity})\times(\text{\# affected})\times (\text{\# susceptible}) .
%\end{equation*}
These assumptions entail the cluster risk $\mu_k = \alpha(n-k) + \beta k(n-k)$.  The susceptible-1 model results from $\beta=0$, and infectivity-1 model is obtained by setting $\alpha=0$.  Testing whether the outcome (positive disease status) clusters in families is equivalent to asking whether $\beta$ is nonzero.  Finding $\beta>0$ might indicate a genetic or household component to disease risk.  The parameterization separates the effect of per-susceiptible risk ($\alpha$) from within-cluster infectivity ($\beta$).  In regression analyses, it is possible to assess how much of the infectivity is due to cluster-level covariates, as we show below in Section \ref{sec:toxicity}.  

\subsubsection{Regression and relative risk for the combined model}

\label{sec:regression}

Suppose we observe $N$ clusters, where $n_i$ is the number of units in cluster $i$ and $r_i$ is the number of affected units in cluster $i$.  In the $i$th cluster, we model the counting process rate as $\mu_k = \alpha_i(n_i-k) + \beta_i k(n_i-k)$ for $k=0,\ldots,n_i$.  Let $d_i$ be a covariate for the $i$th cluster and let $\phi=(\phi_0,\phi_1)$ and $\psi=(\psi_0,\psi_1)$ be covariates.  In toxicology experiments, $d_i$ might correspond to the dose of toxin received by units in cluster $i$.  We use a log-linear parameterization for the counting process rates,  $\log \alpha_i = \phi_0 + \phi_1 d_i$ and $\log\beta_i = \psi_0 + \psi_1 d_i$.  We employ a gradient ascent EM algorithm derived in the Supplementary Appendix to estimate the parameters and standard errors in regression models.

The combined regression model offers an appealing benefit related to the interpretation of risk.  Suppose we estimate $\alpha$ and $\beta$ as in Section \ref{sec:toxicity} under different levels of a dose/exposure $d$ for clustered units.  Then a natural comparison of dose-dependent risk that controls for infectivity of the outcome is the ratio of the per-susceptible risks $\alpha$,
$ \text{RR} =  \left. e^{\phi_0 + \phi_1 d} \middle/ e^{\phi_0} \right. = e^{\phi_1 d}$ .
This is an analogue of the \emph{relative risk} often reported in epidemiological studies under the binomial or Poisson models \citep{McNutt2003Estimating,Zou2004Modified}.  The difference is that RR controls for risk attributable to the interaction of already affected units with susceptible units -- infectivity.  We apply this regression approach in Section \ref{sec:toxicity}.  

%%%%%%%%%%%%%%%%%%%%%%%%%%%%%%%%%%%%%%%%%%%%%%%%%%%

\subsection{The connection}

\label{sec:connection}

Now we show how to construct a sequence of exchangeable dependent Bernoulli variables from a Markov counting process.  The Bernoulli trials are ``embedded'' in the counting process in the following way using probabilistic arguments introduced by \citet{Blom1991Embedding} and \citet[][page 186]{Blom1994Problems}.  To each Bernoulli variable $Z_i$ we associate a latent value $T_i$.  If $\sum_{j=0}^i T_j<t$, where $t>0$ has been chosen in advance, then $Z_i=1$ and otherwise 0.  The $T_i$'s are shown to be equivalent to exponential waiting times in a Markov counting process.
%We emphasize that this is a probabilistic construction of equivalence between two random variables, $X(t)$ and $Y_n$, and not an inference procedure, which we develop in a Supplementary Appendix.  
The relationship between the counting process rates $\mu_k$ and the joint probabilities $\lambda_j$ in the model of \citet{George1995Full} is derived.

%\subsection{Construction of the equivalence}

Consider a set of $n$ units and fix $t>0$ and $\mu_k>0$ for $k=0,\ldots,n-1$ with $\mu_n=0$.  Label the binary response of the $i$th unit $Z_i$.  We construct the responses in $n$ steps.  Let $S_0 = \{1,\ldots,n\}$ represent the indices of the $n$ units initially at risk. \\
\noindent Step 1: For each $i\in S_0$, let $W_i\sim\text{Exponential}(\mu_0/n)$ independently and $T_0=\min\{W_i;\ i\in S_0\}$.  Let $i_1^*$ be the index that achieves this minimum.  Let $Z_{i_1^*}=\indicator{T_0<t}$ and $S_1 = S_0 \setminus \{i_1^*\}$. \\[1em]
Step $k$: For each $i\in S_{k-1}$, let $W_i\sim\text{Exponential}\big(\mu_{k-1}/(n-k+1)\big)$ independently and $T_{k-1} = \min\{W_i,\ i\in S_{k-1}\}$. Let $i_k^*$ be the index that achieves this minimum.  Let $Z_{i_k^*}=\indicator{\sum_{j=0}^{k-1} T_j<t}$, and $S_k = S_{k-1}\setminus \{i_k^*\}$. \\[1em]
Step $n$:  Now $S_{n-1}$ has only one element.  Let $T_{n-1}\sim\text{Exponential}(\mu_{n-1})$ and let $i_n^*$ be the remaining unit.  Let $Z_{i_n^*}=\indicator{\sum_{j=0}^{n-1} T_j < t}$.  \\[1em]
%\end{tabular}
This procedure produces a set of $n$ exchangeable Bernoulli variables $Z_1,\ldots,Z_n$ whose joint probability is given by the transition probability of a counting process.  To see why this is so, recall that since the $W_i$'s at each step are independent, their minimum has exponential distribution with rate equal to the sum of the rates of the $W_i$'s.  At step $k$ we have $T_{k-1} = \min\{W_i;\ i\in S_{k-1}\}$. Since the $W_i$'s are independent, it follows that $T_{k-1} \sim \text{Exponential}\left(\sum_{j=0}^{n-k} \frac{\mu_{k-1}}{n-k+1}\right)= \text{Exponential}(\mu_{k-1})$.

Now consider a Markov counting process $X(t)$ starting at $X(0)=0$.  We can interpret $T_{k-1}$ as the dwell time of the counting process in state $k-1$ before jumping to $k$, so $\sum_{j=0}^{k-1} T_j$ is the time at which the process jumps to state $k$.  Then the probability of $r$ successes is
\begin{equation*}
  \begin{split}
    \Pr\left(Y_n=r\right) &= \Pr(Z_{i_1^*}=\cdots=Z_{i_r^*}=1,\ Z_{i_{r+1}^*}=\cdots=Z_{i_n^*}=0)\\
                        &= \Pr\left(\textstyle\sum_{j=0}^{r-1} T_j < t,\ \sum_{j=0}^{r} T_j > t\right) \\
                        %&= \Pr\big(X(t)=r \mid X(0)=0\big) \\
                        &= P_{0r}(t)
   \end{split}
   \label{eq:equivalence}
 \end{equation*}
 by construction.  In the second line of \eqref{eq:equivalence}, we have replaced the Bernoulli variables $Z_j$ by their corresponding latent variables $T_j$.  In the third line, we have replaced the statements about the sum of waiting times with equivalent statements about the value of the corresponding Markov process $X(t)$ at time $t$.  

To show that the $Z_i$'s thus defined are exchangeable, it suffices to demonstrate that the index $i_k^*$ at each step is chosen uniformly at random from the elements of $S_k$.  We appeal to the notion of competing risks: the waiting time $\min\{W_i;\ i\in S_k\}$ is independent of the particular index $i_k^*$ that achieves this minimum \citep[][page 188]{Lange2010Applied}.  Therefore the probability of choosing any particular $i_k^*$ is given by
$\Pr(i_k^*) = \frac{\mu_k}{n-k} / \sum_{j=1}^{n-k} \frac{\mu_k}{n-k} = \frac{1}{n-k}$.
Then the probability of any particular sequence is $\Pr(i_1^*,\ldots,i_n^*)= 1/n!$ and so the $i_k^*$'s constitute a random permutation of the integers $1,\ldots,n$.  It follows that the count $X(t)$ corresponds to a sum of exchangeable Bernoulli variables.  
%In fact, a Markov counting process can reproduce any probability distribution $\pi_k$ on $\{0,\ldots,n\}$ provided that $\pi_k>0$ for $k\in\{0,\ldots,n\}$ \citep{Faddy1997Extended}.
We emphasize that the times $T_k$ in the counting process representation are auxiliary variables whose purpose is to aid in construction of the equivalence.  It is not necessary to consider $T_k$ to be the waiting time until infection of the $(k+1)$th individual in a familial disease model.  By exchangeability, the order in which the subjects attained their response is irrelevant.  Likewise, the time $t$ is meaningless since scaling $t$ by a constant $c$ and dividing each $\mu_k$ by $c$ does not alter the transition probability.  We henceforth set $t=1$ and write the counting process probability as $P_{0r} = P_{0r}(1)$.

\subsubsection{The relationship between $\mu_k$ and $\lambda_k$ in the \citet{George1995Full} model} 

\label{app:relationship}

The joint success probabilities $\lambda_k$ in the model of \citet{George1995Full} can be derived recursively from the counting process transition probabilities, which are functions of the arrival rates $\mu_k$.  First, note that the probability of $n$ successes in $n$ exchangeable Bernoulli trials is given by
$  \Pr(Y_n=n) = \lambda_n = P_{0n}$
in the counting process model.  Likewise, the probability of $n-1$ successes is given by
$  \Pr(Y_n=n-1) = n[\lambda_{n-1} - \lambda_n] = P_{0,n-1}$ .
Rearranging, we find that 
$  \lambda_{n-1} = \frac{1}{n}P_{0,n-1} + \lambda_n $,
and so on until we reach 
%\begin{equation*}
  %\Pr(Y_n=0) = \sum_{j=0}^n (-1)^j \binom{n}{j}\lambda_j 
%\end{equation*}
%which gives
$  \lambda_0 = P_{00} -  \sum_{j=1}^n (-1)^j \binom{n}{j}\lambda_j$,
recovering each joint probability $\lambda_k$ from the collection of arrival rates in the counting process representation.  Unlike the formulation of \citet{George1995Full}, in which the relationships between the $\lambda_k$'s is complicated, there are no conditions on the rates $\mu_k$ in the Markov process, other than positivity: when all $\mu_k>0$ for $k=0,\ldots,n-1$ and $\mu_n=0$, $P_{mr}$ is always a valid probability distribution on $r\in\{m,\ldots,n\}$.

%\subsection{Simulation of exchangeable Bernoulli sums via counting processes}
%\label{sec:sim1}
%As with other formulations of exchangeable Bernoulli variables, it is always possible to compute the probabilities $\Pr(Y_n=r)$ for $r=0,\ldots,n$ in advance and simulate responses by multinomial sampling on $\{0,\ldots,n\}$ with the probabilities as the sampling weights.  However, the Markov counting process formulation allows direct sampling of outcomes.  Recall that the waiting time $T_k$ between arrival $k$ and $k+1$ has Exponential($\mu_k$) distribution.  The response $Y_n=r$ is the maximum number of arrivals such that $\sum_{j=0}^{r-1} T_j < 1$. Therefore, one simulates exponentially distributed waiting times with rates $\mu_k$ until the sum of waiting times is greater than 1.

\subsection{Ascertainment and different cluster sizes}

\label{sec:ascertainment}

The counting process framework can accommodate data in which clusters are only observed if they meet some condition on the outcome of interest.  For example, in some observational epidemiological studies, only families with one or more affected children are available for study.  When observation is conditional on the outcome of interest, \emph{ascertainment bias} may result.  If only families with $m$ affected members can be studied, the probability of $r$ affected members in a family of size $n$ must be evaluated conditional on having at least $m$ affected members, $P_{mr}$.  In the same way, we can account for clusters of different sizes.  Let $n_i$ be the size of the $i$th cluster and let $r_i$ be the number of units affected.  By specifying the dependence of $\mu_k$ on $n_i$ for $k=0,\ldots,n_i-1$ and letting $\mu_{n_i}=0$, the relevant likelihood is $P_{0r_i}$, evaluated using rates $\mu_k$ that depend on $n_i$.  This is an improvement over previous models, which have generally required that either all clusters be of the same size or that one assume marginal compatibility \citep{Pang2007Test}.

\section{Applications}
\label{sec:applications}

The Supplementary Appendix shows validation results obtained by fitting the proposed models to simulated data.
In this Section, we analyze four datasets that appear to exhibit clustering of responses and compare our results to those obtained using other models, with emphasis on interpretation of estimated parameters.  
%The first three examples deal with aggregation of outcomes in familial studies of disease or death.  In the fourth example, we propose and fit a regression model to data from a developmental toxicology experiment.  
In each case, we compare our results to previous studies using several goodness-of-fit summaries: maximum log-likelihood value ($L$), Akaike information criterion (AIC), Bayesian information criterion (BIC), and $\chi^2$ statistic.  
%We assess the scientific conclusions and interpretations suggested by the fitted models.  
In addition to the standard binomial model, we analyze each dataset using several other models that have shown good performance in previous research on dependent count outcomes:
the beta-binomial model \citep{Moore2001Exploring}, which models overdispersion with respect to binomial outcomes; 
the \citet{Altham1978Two} model for positive and negative association between outcomes; 
the $q$-power model, introduced by \citet{Kuk2004Litter};
the shared response model of \citet{Pang2005Shared} in which a random subset of responses in each cluster are shared; 
the family history (FH) model of \citet{Yu2002Sums} in which the first positive outcome happens with a different probability than subsequent outcomes;
and the incremental risk (IR) model of \citet{Yu2002Sums}.
However, we caution against direct comparison of summaries based on the maximum likelihood value -- the fitted models are quite different and the AIC and BIC may not be suitable for comparison between non-nested models \citep{Dang2009Unified}. 

%Overall, the proposed counting process models provide a superior fit to the data datasets analyzed.
%EM algorithms for maximum likelihood estimation routines are outlined in the Appendix; we use the supplemented EM (SEM) algorithm to compute standard errors for the counting process models \citep{Meng1991Using}. 

\subsection{IPF in families with COPD}

%\begin{table}
  %\centering
  %\begin{tabular}{llccccccc}
%\hline
%Number of & Number of & \multicolumn{7}{l}{Number of } \\
 %siblings & families & \multicolumn{7}{l}{affected siblings }\\
%\hline
%$n$ &     & 0 & 1 & 2 & 3 & 4 & 5 & 6 \\
%\hline
%1 & 48 & 36 & 12 \\
%2 & 23 & 15 & 7 & 1 \\
%3 & 17 & 5  & 7 & 3 & 2 \\
%4 & 7  & 3  & 3 & 1 & 0 & 0 \\
%6 & 5  & 1  & 0 & 1 & 1 & 1 & 0 & 1 \\
%\hline
  %\end{tabular}
  %\caption{Observed frequencies of interstitial pulmonary fibrosis (IPF) in siblings of patients with chronic obstructive pulmonary disease (COPD), from a dataset analyzed by \citet{Liang1992Multivariate} and later by \citet{Yu2002Statistical}.  

\citet{Liang1992Multivariate} present observed frequencies of 60 cases of interstitial pulmonary fibrosis (IPF) in the siblings of families with at least one case of chronic obstructive pulmonary disease (COPD).  %The binomial distribution provides a poor fit to these data, possibly indicating a household or genetic component to disease risk.  
Table \ref{tab:ipffit} presents results.
The FH and IR models of \citet{Yu2002Sums} show good performance in the likelihood based measures ($L$, AIC, and BIC).  The $q$-power and combined models are superior in their $\chi^2$ statistics, with the combined model achieving the lowest value.  
%In the combined model, $\alpha$ has an interpretation as exogenous per-unaffected-unit risk and $\beta$ is the infectivity per contact between affected and unaffected.  Since $\beta$ is significantly nonzero, there may be substantial familial clustering of IPF cases, and a possible household or genetic risk component.  We do not fit the infectivity model to the IPF or Brazilian death data since it requires ascertainment of at least one affected individual.
The Binomial, Beta-binomial, Altham, $q$-power, and shared response models all indicate that the marginal probability of IPF in a single sibling is around 0.3 (the first estimated parameter in the $q$-power model is the marginal probability of ``failure'' -- no IPF).  Each of these indicates positive correlation of IPF cases within families.  Under the FH model, the first affected sibling occurs with low probability, and subsequent siblings are affected with much greater probability.  In the IR model, the risk to unaffected siblings increases monotonically with the number of affected siblings; while baseline risk of IPF is low, each affected sibling substantially increases risk to unaffected siblings.  The Susceptiblle-1 and 2 models show moderate positive association of IPF cases.  The Combined model separates the marginal per-unaffected risk $\alpha$ from the per-contact infectivity $\beta$, indicating substantial contributions of risk from each.

\begin{table}
\label{tab:ipffit}
\centering
\begin{tabular}{llcccccc}
  \hline
  Model    &     & Estimate & SE & $L$ & AIC & BIC & $\chi^2$ \\ 
  \hline
  Binomial & $p$ & 0.296 & 0.032 & -93.0 & 188.1 & 191.4 & 312.3 \\ 
 Beta-Binomial & $p$ & 0.238 & 0.031 & -101.6 & 207.3 & 213.9 & 220.7 \\ 
               & $a$ & 0.086 & 0.057 &  &  &  &  \\ 
        Altham & $p$ & 0.334 & 0.037 & -91.3 & 186.5 & 193.1 & 49.4 \\ 
               & $\theta$  & 0.793 & 0.093 &  &  &  &  \\ 
     $q$-power & $q$ & 0.720 & 0.036 & -87.9 & 179.9 & 186.5 & 12.0 \\ 
               & $\gamma$ & 0.835 & 0.087 &  &  &  &  \\ 
        Shared & $p$ & 0.282 & 0.036 & -89.0 & 182.0 & 188.6 & 21.8 \\ 
               & $\pi$ & 0.439 & 0.098 &  &  &  &  \\ 
            FH & $p$ & 0.177 & 0.032 & -24.0 & 52.0 & 58.6 & 52.1 \\ 
               & $p'$ & 0.549 & 0.111 &  &  &  &  \\ 
            IR & $a$ & -1.533 & 0.215 & -22.1 & 48.1 & 54.8 & 32.2 \\ 
               & $b$ & 1.222 & 0.414 &  &  &  &  \\ 
  \hline
  Susceptible-1 & $\alpha$ & 0.350 & 0.045 & -93.0 & 188.1 & 191.4 & 312.3 \\ 
  Susceptible-2 & $\alpha$ & 0.308 & 0.071 & -92.8 & 189.6 & 196.2 & 258.1 \\ 
                & $\gamma$ & 1.163 & 0.233 &  &  &  &  \\ 
       Combined & $\alpha$ & 0.275 & 0.044 & -87.4 & 178.8 & 185.4 & 9.6 \\ 
                & $\beta$  & 0.300 & 0.124 &  &  &  &  \\ 
   \hline
\end{tabular}
\caption{Results for the IPF dataset.} %  Parameter estimates, standard errors, maximum log-likelihood value ($L$), Akaike information criterion (AIC), Bayesian information criterion (BIC), and $\chi^2$ statistic for each model are given. }
\end{table}

\subsection{Childhood Cancer Syndrome}

\label{sec:cancer}

\citet{Li1988Cancer} report the incidence of cancer in siblings of childhood cancer victims with Li-Fraumeni syndrome from a review of the Cancer Family Registry.  \citet{Yu2002Statistical} present a summary of the data consisting of counts of siblings of children with cancer.  In our analysis, we account for ascertainment of families via a single affected child by the conditioning argument outlined in Section \ref{sec:ascertainment}.  Therefore, the dataset we analyze here is the same as that presented in \citet{Yu2002Statistical}, but adjusted to include the affected children.  Table \ref{tab:cancer} shows the results.  The IR, Susceptible, and Combined models achieve the best likelihood-based scores, with the Infective-2 and Susceptible-2 models having the lowest $\chi^2$ value.  The first models in Table \ref{tab:cancer} indicate that the marginal probability of childhood cancer in already-affected families is large, between 0.4 and 0.5.  There may be correlation in the outcomes of individuals in these families, but the considered models disagree about its sign.  The Beta-Binomial, $q$-power, and IR models indicate negative correlation, but the Altham model (and the Shared Response model, by design)  indicates positive association. The Infective, Susceptible, and Combined models offer an alternative explanation: each affected sibling increases the risk to others, but this increase diminishes as more siblings are affected.  Notably, there is little evidence from these models of increased per-contact risk due to infectivity.  We do not fit the FH model of \citet{Yu2002Sums} to the cancer dataset since only families with one affected child were ascertained.

\begin{table}
  \centering
\begin{tabular}{llcccccc}
  \hline
  Model & & Estimate & SE & $L$ & AIC & BIC & $\chi^2$ \\ 
  \hline
  Binomial & $p$ & 0.487 & 0.047 & -34.5 & 71.1 & 73.8 & 39.5 \\ 
 Beta-Binomial & $p$ & 0.436 & 0.043 & -40.5 & 85.0 & 90.5 & 99.6 \\ 
               & $a$ & -0.043 & 0.046 &  &  &  &  \\ 
        Altham & $p$ & 0.488 & 0.045 & -34.5 & 73.0 & 78.5 & 38.5 \\ 
               & $\theta$ & 0.970 & 0.105 &  &  &  &  \\ 
     $q$-power & $q$ & 0.493 & 0.058 & -33.9 & 71.9 & 77.3 & 35.6 \\ 
               & $\gamma$ & 0.911 & 0.088 &  &  &  &  \\ 
        Shared & $p$ & 0.494 & 0.059 & -34.5 & 73.1 & 78.5 & 39.0 \\ 
               & $\pi$ & 0.135 & 0.325 &  &  &  &  \\ 
            IR & $a$ & 1.403 & 0.624 & -27.5 & 59.0 & 64.5 & 37.6 \\ 
               & $b$ & -1.132 & 0.390 &  &  &  &  \\ 
  \hline
  Infective-1 & $\beta$ & 0.275 & 0.051 & -35.1 & 72.2 & 74.9 & 45.3 \\ 
  Infective-2 & $\beta$ & 0.739 & 0.222 & -27.8 & 61.5 & 69.7 & 22.9 \\ 
              & $\eta$  & $<0.001$ & 0.536 &  &  &  &  \\ 
              & $\gamma$ & 0.434 & 0.246 &  &  &  &  \\ 
Susceptible-1 & $\alpha$ & 0.428 & 0.078 & -29.5 & 60.9 & 63.7 & 29.6 \\ 
Susceptible-2 & $\alpha$ & 0.904 & 0.321 & -27.2 & 58.4 & 63.8 & 21.5 \\ 
              & $\gamma$ & 0.433 & 0.257 &  &  &  &  \\ 
     Combined & $\alpha$ & 0.384 & 0.219 & -29.7 & 63.3 & 68.8 & 31.0 \\ 
              & $\beta$ & $<0.001$ & 0.139 &  &  &  &  \\ 
  \hline
\end{tabular}
\caption{Results for the childhood cancer data. }
\label{tab:cancer}
\end{table}

\subsection{Childhood Mortality in Brazil}

\label{sec:deaths}

\citet{Yu2002Statistical} summarize data first reported by \citet{Sastry1997Nested} on deaths of children in families of various sizes in a study of childhood mortality in impoverished areas of Brazil.  \citet{Yu2002Statistical} note that family size appears to correlate with mortality and show that the FH and IR models fit the data well.  Table \ref{tab:deaths} gives the results, with the FH and IR models showing the best likelihood-based measures, and the combined model clearly outperforming the others in its $\chi^2$ statistic.  The marginal probability of death of a single child is estimated to be slightly larger than 0.1 in this population, and the correlation of responses is estimated by most models to be positive, with the exception of the Altham model, where $\theta>1$; the large standard error and $\chi^2$ value here suggest that the Altham model fits these data poorly.  The Susceptible models offer little insight, but the Combined model tells a fuller story: baseline risk to a given child is low, but the risk to the family depends both on the number of children who have died, and the number remaining.  This suggests that the childhood mortality may have a ``contagious'' component within families in this community.

\begin{table}
  \centering
\begin{tabular}{llcccccc}
  \hline
  Model & & Estimate & SE & $L$ & AIC & BIC & $\chi^2$ \\ 
  \hline
  Binomial & $p$ & 0.146 & 0.007 & -791.9 & 1585.7 & 1591.7 & 2300.4 \\ 
 Beta-Binomial & $p$ & 0.134 & 0.007 & -773.1 & 1550.1 & 1562.1 & 135.5 \\ 
               & $a$ & 0.115 & 0.023 &  &  &  &  \\ 
        Altham & $p$ & 0.123 & 0.010 & -788.0 & 1579.9 & 1591.9 & 8788.0 \\ 
               & $\theta$ & 1.105 & 0.040 &  &  &  &  \\ 
     $q$-power & $q$ & 0.859 & 0.007 & -774.3 & 1552.7 & 1564.7 & 135.5 \\ 
               & $\gamma$ & 0.915 & 0.023 &  &  &  &  \\ 
        Shared & $p$ & 0.137 & 0.007 & -766.6 & 1537.2 & 1549.2 & 124.6 \\ 
               & $\pi$ & 0.323 & 0.031 &  &  &  &  \\ 
            FH & $p$ & 0.111 & 0.007 & -459.2 & 922.5 & 934.5 & 338.9 \\ 
               & $p'$ & 0.300 & 0.024 &  &  &  &  \\ 
            IR & $a$ & -2.043 & 0.064 & -458.8 & 921.6 & 933.6 & 271.2 \\ 
               & $b$ & 0.813 & 0.101 &  &  &  &  \\ 
  \hline
  Susceptible-1 & $\alpha$ & 0.158 & 0.008 & -791.9 & 1585.7 & 1591.7 & 2300.5 \\ 
  Susceptible-2 & $\alpha$ & 0.066 & 0.010 & -764.9 & 1533.8 & 1545.7 & 3847.2 \\ 
                & $\gamma$ & 1.716 & 0.104 &  &  &  &  \\ 
       Combined & $\alpha$ & 0.123 & 0.007 & -750.3 & 1504.6 & 1516.6 & 67.4 \\ 
                & $\beta$ & 0.159 & 0.023 &  &  &  &  \\ 
   \hline
\end{tabular}
\caption{Results for the Brazilian childhood mortality data.  }
\label{tab:deaths}
\end{table}

\subsection{Developmental toxicity of an herbicide}

\label{sec:toxicity}

Researchers exposed pregnant mice to different doses of the herbicide 2,4,5-trichlorophenoxyacetic acid (2,4,5-T) during gestation and recorded the number of implanted fetuses and the number of fetuses that died, were resorbed, or had a cleft palate \citep{Holson1992Developmental,Chen1992Correlations}.  They observed the number of implanted fetuses, number of ``affected'' fetuses, and the dose of 2,4,5-T for each mouse in the experiment and are given in Table 1 of \citet{George1995Full}.  The mice were grouped into six levels, receiving doses of 0, 30, 45, 60, 75, or 90 mg/kg of 2,4,5-T.  The responses of litter-mates are correlated because the fetuses gestate in the same mother. 
%and researchers wish to estimate the dose-dependent toxicity of 2,4,5-T.  
Let $n_i$ be the number of implanted fetuses (cluster size) in dam $i$, let $d_i$ be the dose, and let $r_i$ be the number of fetuses affected.  We fit the Combined model with covariate vector $z_i=(1,d_i)$.
%Then letting $z_i=(1,d_i)$ be the cluster-specific covariate vector for the $i$th observation, we form the combined model as follows.  

The results of the regression are given in Table \ref{tab:toxicity}.  The first two lines give estimates and standard errors for the elements of $\phi$ and $\psi$.  The next lines give $\alpha$ and $\beta$, stratified by different dose level, where the standard errors were obtained by the delta method.  Both $\alpha$ and $\beta$ increase with dose level, and $\beta$ increases much more quickly than $\alpha$.  Therefore both exogenous risk and within-cluster effects appear to be significantly related to the number of affected fetuses -- and litter size -- in this experiment.  The baseline risk and infectivity are very small in the absence of 2,4,5-T, and the ``infectivity'' of each affected fetus increases with dose.  We obtain $L = -753.0$ and $\chi^2 = 1044.4$ for the fitted model.  In toxicity trials, the relationship between dose and risk for individual units is often of greatest interest.  
%Many phenomenological models for the outcome (e.g.  Binomial, Beta-binomial, Altham, $q$-power, Shared response) provide the marginal probability of one unit being affected, across all possible outcomes for other units.  But the possibility of infectivity or interaction between affected and unaffected units can make interpretation of this marginal probability difficult.  Higher doses of toxin may result in more affected units, which may increase risk purely due to infectivity.  
%The Combined model in this example gives a simple way to estimate the marginal per-susceptible risk $\alpha$ due to toxicity, while controlling for infectivity or interaction between affected and unaffected units.  
%This relationship is summarized by the relative risk (RR) introduced in Section \ref{sec:regression}.  
Letting $d$ be the dose of toxin delivered, $e^{\hat{\phi}_1 d}$ is an estimate of the dose-dependent relative risk to unaffected units, not due to contagion.  Table \ref{tab:rr} gives estimates and standard errors for the RR in this experiment.  For example, at dose 90 mg/kg, 2,4,5-T delivers a more than four-fold increase in the risk to an individual fetus, over that to which a fetus gestating in a control ($d=0$) mouse is subject.

%\citet{Kuk2004Litter} find that fitting the beta-binomial model gives $L=-736.2$ and the $q$-power model gives $L=-702.8$.

% \citet{Dang2009Unified} fit several models to a modified form of this dataset without the 90mg/kg dose level.  

%For this modified dataset, we find that $L=$ and $\chi^2=$ [need to re-fit]

%[comparison to other fits of the same data? See Table V of Dang (2009). ]

%To determine a safe dose level, \citet{Kuk2004Litter} and \citet{Dang2009Unified} use the probability $p$ that there is at least one affected fetus.  For the counting process models introduced here, this probability is always function of cluster (litter) size $n$, and we denote it $p_n$.  Then
%\[ p_n = \Pr(Y_n \geq 1) = 1-\Pr(Y_n=0). \] 
%At a particular dose level, we denote this probability $p_n(d)$.  Following \citet{Kuk2004Litter}, let $f(n)$ be the frequency of litter size $n$ and let
%\[ r(d) = \sum_{n=1}^\infty f(n) p_n(d) \]
%be the dose-specific marginal probability that at least one fetus is affected, across all litter sizes $n$.  Then $r^*(d) = r(d) - r(0)$ is the excess risk at dose $d$ over the ``background risk'' at dose 0, $r(0)$.  We first seek a benchmark dose: the largest dose $d$ that gives $r^*(d) < \alpha$. 

%Using the maximum likelihood estimates given in Table \ref{tab:toxicity}, we find that 

%safe dose level:
%(Crump, 1984; Chen and Kodell, 1989; Ryan, 1992)

\begin{table}
  \begin{center}
  \begin{tabular}{cccccccccc}
\hline
Dose    && \multicolumn{3}{c}{Extra-Cluster risk} &&& \multicolumn{3}{c}{Infectivity} \\ \cline{3-5} \cline{8-10}
(mg/kg) && Parameter & Estimate & SE    &&& Parameter & Estimate & SE        \\
\hline
All && $\phi_0$                    & -2.760 & 0.122 &&& $\psi_0$  & -3.453   &  0.177    \\
    && $\phi_1$                     & 0.016 & 0.003 &&& $\psi_1$  & 0.042    &  0.003    \\[1em]
0   && $\alpha=e^{\phi_0}$          & 0.063 & 0.122 &&& $\beta=e^{\psi_0}$          & 0.032 & 0.177 \\[1em]
30 && $\alpha=e^{\phi_0+30\phi_1}$  & 0.103 & 0.144 &&& $\beta=e^{\psi_0+30\psi_1}$ & 0.113 & 0.203 \\[1em]
45 && $\alpha=e^{\phi_0+45\phi_1}$  & 0.132 & 0.168 &&& $\beta=e^{\psi_0+45\psi_1}$ & 0.214 & 0.231 \\[1em]
60 && $\alpha=e^{\phi_0+60\phi_1}$  & 0.168 & 0.196 &&& $\beta=e^{\psi_0+60\psi_1}$ & 0.404  & 0.265 \\[1em]
75 && $\alpha=e^{\phi_0+75\phi_1}$  & 0.214 & 0.227 &&& $\beta=e^{\psi_0+75\psi_1}$ & 0.764 & 0.304 \\[1em]
90 && $\alpha=e^{\phi_0+90\phi_1}$  & 0.273 & 0.260 &&& $\beta=e^{\psi_0+90\psi_1}$ & 1.444 & 0.345 \\
\hline
\end{tabular}
\end{center}
\caption{Combined model regression estimates and standard errors for the developmental toxicity data in Table 1 of \citet{George1995Full}.  The overall results for the parameters $\phi_0$, $\phi_1$, $\psi_0$, and $\psi_1$ are given in the first two lines.  Below, exogenous risk ($\alpha$) and infectivity ($\beta$) parameters are given for each dose level, where $\alpha=\exp[z_i' \phi]$, $\beta=\exp[z_i'\psi]$ and $z_i=(1, \text{dose})$.  Standard errors of $\alpha$ and $\beta$ for the different dose levels were obtained by the delta method.}
\label{tab:toxicity}
\end{table}

\begin{table}
  \begin{center}
    \begin{tabular}{lccc}
      \hline
      Dose & Expression     & RR & SE \\
      \hline
      0          & $e^0$          & 1  &    \\
      30         & $e^{30\phi_1}$ & 1.628 & 0.0178 \\
      45         & $e^{45\phi_1}$ & 2.078 & 0.0246 \\
      60         & $e^{60\phi_1}$ & 2.652 & 0.0321 \\
      75         & $e^{75\phi_1}$ & 3.384 & 0.0405 \\
      90         & $e^{90\phi_1}$ & 4.318 & 0.0502 \\
      \hline
  \end{tabular}
\end{center}
\caption{Relative risk (RR) estimates and standard errors for the Combined model in the developmental toxicology data.  Standard errors were obtained by the delta method.}
\label{tab:rr}
\end{table}

%%%%%%%%%%%%%%%%%%%%%%%%%%%%%%%%%%%%%%%%%%%
\section{Discussion}

The paradigm of \citet{George1995Full} is useful because the likelihood for any dependency model of exchangeable Bernoulli variables can be expressed simply.  However, it can be difficult to translate knowledge of the dependency pattern into the joint outcome probabilities necessary to write the likelihood.  
%Models that are flexible enough to fit a variety of datasets well can be difficult to interpret.  
%The assumption of marginal compatibility presented by \citet{Xu2003Modelling} and \citet{Pang2007Test} can assist in interpretation of the joint probabilities $\lambda_j$ in these models.  But this assumption implicitly disallows models that depend on cluster size $n$.  There has also been uncertainty about how to incorporate data from different cluster sizes; joint probabilities and higher-order correlations are not the same in general when the number of Bernoulli trials changes \citep{Stefanescu2003Likelihood,Matthews2005Analysis}. Often clusters are identified through one or more subjects satisfying certain criteria related to their response, creating a possible bias in estimation results \citep{Wickramaratne2004Approaches,Matthews2008Analysis}.  
In this work, we have developed a flexible class of Markov counting models for analyzing clustered binary data.  We have established a correspondence between these models and sums of dependent Bernoulli variables under the framework of \citet{George1995Full}.  We believe the combined model outlined in section \ref{sec:si} is most useful.  Inference under this model addresses a fundamental question in infectious disease epidemiology: estimating $\beta>0$ means that some disease risk is due to infectivity or interaction between affected or unaffected units in a cluster.

\noindent \textbf{Supplementary Materials:} In a Supplementary Appendix, we outline maximum likelihood estimation via the EM algorithm, regression for the combined model, and a method for numerically evaluating counting process likelihoods.\\
\textbf{Acknowledgements:} We thank Theodore R. Holford and Hongyu Zhao for helpful comments on the manuscript.

%%%%%%%%%%%%%%%%%%%%%%%%%%%%%%%%%%%%%%%%%%%%%%%%%
% Bibliography 
\bibliographystyle{biorefs}
\bibliography{dep}
%%%%%%%%%%%%%%%%%%%%%%%%%%%%%%%%%%%%%%%%%%%%

\end{document}